\documentstyle[twoside]{article}
 
\catcode`\@=11
\long\def\@makefntext#1{
\protect\noindent \hbox to 3.2pt {\hskip-.9pt
$^{{\eightrm\@thefnmark}}$\hfil}#1\hfill}               

\def\@makefnmark{\hbox to 0pt{$^{\@thefnmark}$\hss}}    
 
\def\ps@myheadings{\let\@mkboth\@gobbletwo
\def\@oddhead{\hbox{}
\rightmark\hfil\eightrm\thepage}
\def\@oddfoot{}\def\@evenhead{\eightrm\thepage\hfil
\leftmark\hbox{}}\def\@evenfoot{}
\def\sectionmark##1{}\def\subsectionmark##1{}}
 
\oddsidemargin=\evensidemargin
\newcounter{sectionc}\newcounter{subsectionc}\newcounter{subsubsectionc}
\renewcommand{\section}[1] {\vspace{12pt}\addtocounter{sectionc}{1}
\setcounter{subsectionc}{0}\setcounter{subsubsectionc}{0}\noindent
        {\tenbf\thesectionc. #1}\par\vspace{5pt}}
\renewcommand{\subsection}[1] {\vspace{12pt}\addtocounter{subsectionc}{1}
        \setcounter{subsubsectionc}{0}\noindent
        {\bf\thesectionc.\thesubsectionc. {\kern1pt \bfit #1}}\par\vspace{5pt}}
\renewcommand{\subsubsection}[1] {\vspace{12pt}\addtocounter{subsubsectionc}{1}
        \noindent{\tenrm\thesectionc.\thesubsectionc.\thesubsubsectionc.
        {\kern1pt \tenit #1}}\par\vspace{5pt}}
\newcommand{\nonumsection}[1] {\vspace{12pt}\noindent{\tenbf #1}
        \par\vspace{5pt}}
 
\newcounter{appendixc}
\newcounter{subappendixc}[appendixc]
\newcounter{subsubappendixc}[subappendixc]
\renewcommand{\thesubappendixc}{\Alph{appendixc}.\arabic{subappendixc}}
\renewcommand{\thesubsubappendixc}
        {\Alph{appendixc}.\arabic{subappendixc}.\arabic{subsubappendixc}}
 
\renewcommand{\appendix}[1] {\vspace{12pt}
        \refstepcounter{appendixc}
        \setcounter{figure}{0}
        \setcounter{table}{0}
        \setcounter{lemma}{0}
        \setcounter{theorem}{0}
        \setcounter{corollary}{0}
        \setcounter{definition}{0}
        \setcounter{equation}{0}
        \renewcommand{\thefigure}{\Alph{appendixc}.\arabic{figure}}
        \renewcommand{\thetable}{\Alph{appendixc}.\arabic{table}}
        \renewcommand{\theappendixc}{\Alph{appendixc}}
        \renewcommand{\thelemma}{\Alph{appendixc}.\arabic{lemma}}
        \renewcommand{\thetheorem}{\Alph{appendixc}.\arabic{theorem}}
        \renewcommand{\thedefinition}{\Alph{appendixc}.\arabic{definition}}
        \renewcommand{\thecorollary}{\Alph{appendixc}.\arabic{corollary}}
        \renewcommand{\theequation}{\Alph{appendixc}.\arabic{equation}}
        \noindent{\tenbf Appendix \theappendixc #1}\par\vspace{5pt}}
\newcommand{\subappendix}[1] {\vspace{12pt}
        \refstepcounter{subappendixc}
        \noindent{\bf Appendix \thesubappendixc. {\kern1pt \bfit #1}}
        \par\vspace{5pt}}
\newcommand{\subsubappendix}[1] {\vspace{12pt}
        \refstepcounter{subsubappendixc}
        \noindent{\rm Appendix \thesubsubappendixc. {\kern1pt \tenit #1}}
        \par\vspace{5pt}}
 
\newcommand{\textlineskip}{\baselineskip=13pt}
\newcommand{\smalllineskip}{\baselineskip=10pt}
 
\def\eightcirc{
\begin{picture}(0,0)
\put(4.4,1.8){\circle{6.5}}
\end{picture}}
\def\eightcopyright{\eightcirc\kern2.7pt\hbox{\eightrm c}}
 
\newcommand{\copyrightheading}[1]
        {\vspace*{-2.5cm}\smalllineskip{\flushleft
        {\footnotesize $\eightcopyright$\, World Scientific Publishing
         Company}\\
         }}
 
\def\abstracts#1#2#3{{
        \centering{\begin{minipage}{4.5in}\baselineskip=10pt\footnotesize
        \parindent=0pt #1\par
        \parindent=15pt #2\par
        \parindent=15pt #3
        \end{minipage}}\par}}
 
\def\keywords#1{{
        \centering{\begin{minipage}{4.5in}\baselineskip=10pt\footnotesize
        {\footnotesize\it Keywords}\/: #1
        \end{minipage}}\par}}
 
\renewenvironment{thebibliography}[1]
        {\frenchspacing
         \ninerm\baselineskip=11pt
         \begin{list}{\arabic{enumi}.}
        {\usecounter{enumi}\setlength{\parsep}{0pt}
         \setlength{\leftmargin 12.7pt}{\rightmargin 0pt} 
         \setlength{\itemsep}{0pt} \settowidth
        {\labelwidth}{#1.}\sloppy}}{\end{list}}
 
\newcounter{itemlistc}
\newcounter{romanlistc}
\newcounter{alphlistc}
\newcounter{arabiclistc}

\newcommand{\fcaption}[1]{
        \refstepcounter{figure}
        \setbox\@tempboxa = \hbox{\footnotesize Fig.~\thefigure. #1}
        \ifdim \wd\@tempboxa > 5in
           {\begin{center}
        \parbox{5in}{\footnotesize\smalllineskip Fig.~\thefigure. #1}
            \end{center}}
        \else
             {\begin{center}
             {\footnotesize Fig.~\thefigure. #1}
              \end{center}}
        \fi}
 
\newcommand{\tcaption}[1]{
        \refstepcounter{table}
        \setbox\@tempboxa = \hbox{\footnotesize Table~\thetable. #1}
        \ifdim \wd\@tempboxa > 5in
           {\begin{center}
        \parbox{5in}{\footnotesize\smalllineskip Table~\thetable. #1}
            \end{center}}
        \else
             {\begin{center}
             {\footnotesize Table~\thetable. #1}
              \end{center}}
        \fi}
 
\def\@citex[#1]#2{\if@filesw\immediate\write\@auxout
        {\string\citation{#2}}\fi
\def\@citea{}\@cite{\@for\@citeb:=#2\do
        {\@citea\def\@citea{,}\@ifundefined
        {b@\@citeb}{{\bf ?}\@warning
        {Citation `\@citeb' on page \thepage \space undefined}}
        {\csname b@\@citeb\endcsname}}}{#1}}
 
\newif\if@cghi
\def\cite{\@cghitrue\@ifnextchar [{\@tempswatrue
        \@citex}{\@tempswafalse\@citex[]}}
\def\citelow{\@cghifalse\@ifnextchar [{\@tempswatrue
        \@citex}{\@tempswafalse\@citex[]}}
\def\@cite#1#2{{$\null^{#1}$\if@tempswa\typeout
        {IJCGA warning: optional citation argument
        ignored: `#2'} \fi}}

\def\pmb#1{\setbox0=\hbox{#1}
        \kern-.025em\copy0\kern-\wd0
        \kern.05em\copy0\kern-\wd0
        \kern-.025em\raise.0433em\box0}

\def\fnt#1#2{\footnotetext{\kern-.3em
        {$^{\mbox{\scriptsize #1}}$}{#2}}}
 
\def\runninghead#1#2{\pagestyle{myheadings}
\markboth{{\protect\footnotesize\it{\quad #1}}\hfill}
{\hfill{\protect\footnotesize\it{#2\quad}}}}
\font\tenrm=cmr10
\font\tenit=cmti10
\font\tenbf=cmbx10
\font\bfit=cmbxti10 at 10pt
\font\ninerm=cmr9

\font\eightrm=cmr8

\def\qed{\hbox{${\vcenter{\vbox{                        
   \hrule height 0.4pt\hbox{\vrule width 0.4pt height 6pt
   \kern5pt\vrule width 0.4pt}\hrule height 0.4pt}}}$}}
 
\def\bsc{{\sc a\kern-6.4pt\sc a\kern-6.4pt\sc a}}       
\def\bflatex{\bf L\kern-.30em\raise.3ex\hbox{\bsc}\kern-.14em
T\kern-.1667em\lower.7ex\hbox{E}\kern-.125em X}
 
\input epsf.tex
\begin{document}
\runninghead{
Towards automatic analytic evaluating of massive Feynman diagrams
} {
Towards automatic analytic evaluating of massive Feynman diagrams
}
 
\normalsize\textlineskip
\thispagestyle{empty}
\setcounter{page}{1}
 
\copyrightheading{}                     
 
\vspace*{0.88truein}
 
\centerline{\bf TOWARDS AUTOMATIC ANALYTIC EVALUATION}
\centerline{\bf OF MASSIVE FEYNMAN DIAGRAMS.}
\vspace*{0.035truein}
\centerline{\footnotesize
L.~Avdeev
\footnote{
E-mail: avdeevL@thsun1.jinr.dubna.su
~Supported in part RFFI grant \# 96-02-17531, JSPS FSU Project.
},~
J.~Fleischer
\footnote{
E-mail: fleischer@physik.uni-bielefeld.de
},~
M.~Yu.~Kalmykov 
\footnote{
E-mail: misha@physik.uni-bielefeld.de
~Supported in part by Volkswagenstiftung.
}~
and M.~Tentyukov
\footnote{
E-mail: tentukov@physik.uni-bielefeld.de
~Supported by Bundesministerium f\"ur Forschung und Technologie.
}
}

\vspace*{0.015truein}
\centerline{\footnotesize\it Fakult\"at f\"ur Physik, Universit\"at Bielefeld}
		    
\baselineskip=10pt
\centerline{\footnotesize\it D-33615 Bielefeld 1, Germany}
\vspace*{0.225truein}
 \vspace*{0.21truein}
\abstracts{
A method to calculate two-loop self-energy diagrams of the Standard
Model is demonstrated. A direct physical application is the calculation
of the two-loop electroweak contribution to the anomalous magnetic
moment of the muon
${\frac{1}{2}(g-2)}_{\mu}$. Presently we confine ourselves to a ``toy''
model with only $\mu$, $\gamma$ and a scalar particle (Higgs).
The algorithm is implemented as a package of computer programs in FORM. 
For generating and automatically evaluating any number of two-loop 
self-energy diagrams, a special C-program has been written. This program 
creates the initial FORM-expression for every diagram generated by
QGRAF,  executes the corresponding subroutines
and sums up the final results.
}{}{}
 
\vspace*{10pt}
\vspace*{1pt}\textlineskip      
\section{Introduction}
\vspace*{-0.5pt}
\noindent
   Recent high precision experiments to verify the Standard Model
of electroweak interactions require, on the side of the theory, high
precision calculations resulting in the evaluation of higher loop diagrams. 
For specific processes thousands of 
multi-loop Feynman diagrams do contribute, and it turns out impossible to 
perform these calculations by hand. That makes the request for 
automatization a high priority task. In this direction,
several program packages are elaborated
(\cite{GRACE} \cite{FeynmArts} and reports of the 
Minami-Tateya collaboration, Bauberger et al., Pukhov et al.,  
Vermaseren and J.X.Wang on this conference). It appears absolutely 
necessary that various groups 
produce their own solutions of handling this problem: the various ways
will be of different efficiency, have different domains of applicability,
and last but not least, should eventually allow for completely
independent checks of the final results. This point of view motivated us
to seek our own way for automatic evaluation of Feynman diagrams. We 
have in mind only higher loop calculations (no multipoint functions).

   We demonstrate here the functioning of a C-program (TLAMM) for the
evaluation of the two-loop anomalous magnetic moment (AMM) of the muon 
${\frac{1}{2}(g-2)}_{\mu}$, but
on the same lines we have developed the program to an extent
that it can be applied to arbitrary processes in the Standard Model.
This piloting C-program must read the diagrams generated by QGRAF 
\cite{QGRAF} for a given physical process, generate the FORM \cite{FORM}
source code, start the FORM interpreter, read and sum up the results 
for the class of diagrams under consideration.  For the purpose of
demonstration, here we apply TLAMM to a closed subclass of diagrams of 
the Standard Model, which we refer to as a ``toy'' model.

Recent papers reduced the theoretical uncertainty of the muon AMM 
by partially calculating the two-loop electroweak contributions
\cite{Kuraev}, \cite{Czarnecki}. 
Their results have been obtained in the following approximation:
terms suppressed by $(1-4 \sin^2 \Theta )$ were omitted;
the fermion masses of the first two families are set to zero; in the third
family the $\tau$- and $b$-quark masses are taken to be zero as well;
diagrams with two or more scalar couplings to the muon are
suppressed by the ratio $\frac{m^2_{\mu}}{M^2_Z}$ and have been discarded;
the Kobayashi-Maskawa matrix is assumed to be unity;
the mass of the Higgs particle is large compared to $M_{Z,W}$.

All of these approximations, except possibly the last one, are well
justified and give rise to small corrections only. We consider it of
great interest to study also the case $M_H \sim M_{Z,W}$. 
To perform this calculation is our main physical motivation. Apart from
that, for technical reasons, it may be interesting to study the
functioning of TLAMM by calculating all $1832$ two-loop diagrams 
without any approximation.

The calculation of the anomalous magnetic moment of the muon 
reduces, after differentiation and contractions with projection 
operators, to diagrams of propagator type with external momentum
on the muon mass-shell (for details see \cite{projector}).  

The method of Taylor
expansion of diagrams in external momenta and Pad\'e approximation
\cite{Pade} yields only slowly converging results at the threshold. 
Since in the case of the muon AMM we are confronted with a low
energy problem it appears natural to expand with respect to the heavy
masses ($M_Z, M_H$ and $m_t$) of the theory. 
The applicability of the ``asymptotic-R-operation'' 
\cite{asymptotic} in the limit of large masses 
has to be investigated for diagrams evaluated on the muon
mass-shell(i.e. $p^2=-m_{\mu}^2$). Some diagrams already had a
threshold at the muon mass-shell before the expansion. 
In other diagrams
this threshold appears in some terms of the expansion. 
In dimensional regularization, threshold singularities
(like any other infrared singularities if they are strong enough) 
manifest themselves as poles 
in $\varepsilon$ (d=4-2$\varepsilon$). They ought to cancel for 
the total AMM. We check this in our toy model.

\section{Large-mass expansion}
\noindent
The asymptotic expansion in the limit of large masses is defined 
\cite{Davydychev} as 

\begin{equation}
F_G(q, M ,m, \varepsilon) \stackrel{M \to \infty}{\sim }
\sum_{\gamma} F_{G/\gamma}(q,m,\varepsilon) \circ
T_{q^{\gamma}, m^{\gamma}}
F_{\gamma}(q^{\gamma}, M ,m^{\gamma}, \varepsilon)
\end{equation}
where $G$ is the original graph, $\gamma$'s are subgraphs involved in the
asymptotic expansion, $G/\gamma$ denotes shrinking $\gamma$ to a point;
$F_{\gamma}$ is the Feynman integral corresponding to $\gamma$; 
$ T_{q^{\gamma}, m^{\gamma}} $ is the Taylor operator expanding 
the integrand in small masses $\{ m_{\gamma} \}$ and external momenta
$\{ q_{\gamma} \}$ of the subgraph $\gamma$ (before integration);
``$ \circ $'' inserts the subgraph expansion 
in the numerator of the integrand $F_{G/{\gamma}}$. 
The sum goes over all subgraphs $\gamma$ which
(a) contain all lines with large masses, and
(b) are one-particle irreducible relative to lines with light masses.

The following types of integrals occur in the asymptotic expansion 
of the muon AMM in the Standard Model:
$1.$
two-loop tadpole diagrams with various heavy masses on internal
lines;
$2.$
two-loop self-energy diagrams, involving contributions from fermions
lighter or heavier than the muon, with the external momentum on
the muon mass-shell;
$3.$
two-loop self-energy diagrams with two or three muon  lines and
the external momentum on the muon  mass-shell;
$4.$
products of one-loop self-energy diagrams on-shell and a one-loop tadpole 
with a heavy mass.
Almost all of these diagrams can be evaluated analytically using the package
SHELL2 \cite{SHELL2}. For our calculation we have modified this package
in the following way:
$1.$
There are no restrictions on the indices on the lines
(powers of scalar denominators).
$2.$ 
More recurrence relations are used, and the dependence on the space-time
dimension is always explicitly reducible to powers of linear factors.
$3.$
A new algorithm for simplification of this rational fractions is 
implemented. These modifications 
essentially reduce the execution time (in some cases, down to 
the order of a hundredth).
$4.$
New programs for evaluating two-loop tadpole integrals with different
masses are added. 
$5.$
New programs were written for the asymptotic expansion of one-loop
self-energy diagrams (relevant for renormalization)
in the large-mass limit.

\section{The toy-model}
\noindent
As the first step, we concentrate on a ``toy'' model, a ``slice'' of the 
Standard Model, involving a light charged spinor $\Psi$, the photon $A_\mu$, 
and a heavy neutral scalar field $\Phi$. 
The scalar has triple $\left( g \right)$ and quartic $\left( \lambda \right)$
self-interactions, and the Yukawa coupling to the spinor $\left( y \right)$.
The Lagrangian of the toy-model reads (in the Euclidean space-time):

\begin{eqnarray}
{\it L} & = & \frac{1}{2} \partial_\mu \Phi \partial^\mu \Phi
+ \frac{1}{2} M^2 \Phi^2 - \frac{g}{3!} \Phi^3 - \frac{\lambda}{4!}
\Phi^4
+ \frac{1}{4} F_{\mu \nu}^2 + \frac{1}{2 \alpha}
\left( \partial_\mu A^\mu \right)^2
\nonumber \\
&& + \bar{\Psi} \left( \hat{\partial} + m \right) \Psi
+ i e \bar{\Psi} \hat A \Psi - y \Phi \bar{\Psi} \Psi
\end{eqnarray}
where $e$ is the electric charge and $\alpha$ is a gauge fixing parameter.

The main aims of the present investigation are the following:

$1.$
Verification of the consistency of the large-mass asymptotic expansion 
with the external momentum on the mass shell of a small mass.
In particular, we check the cancelation of all
threshold singularities, that appear in individual diagrams
and manifest themselves as infrared poles in $\varepsilon$.

$2.$
Estimation of the influence of a heavy neutral scalar particle on the AMM
of the muon in the framework of the Standard Model. 

$3.$
Verification of gauge independence: we use the covariant
gauge with an arbitrary parameter $\alpha$.

\begin{figure}[t]
\vskip 80mm
\centerline{\vbox{\epsfysize=35mm \epsfbox{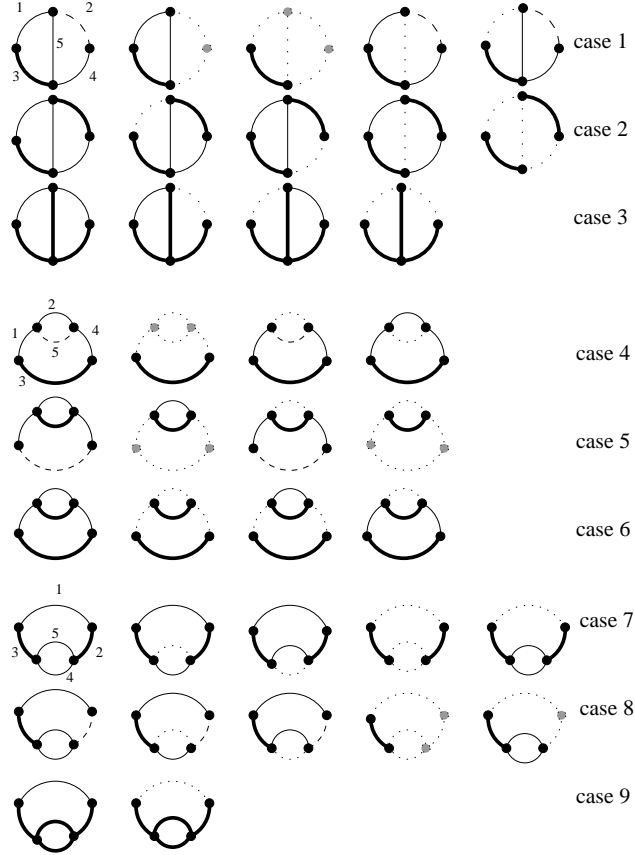}}}
\caption{\label{fig3} The prototypes and their subgraphs
contributing
to the large mass expansion. Bold, thin and dashed lines correspond to
the heavy-mass, light-mass, and massless propagators, respectively.
Dotted lines indicate the lines omitted in the subgraph $\gamma$.}
\end{figure}

In the following we  analyze in some detail the diagrams contributing 
to the AMM of the fermion in our toy-model and specify the 
renormalization procedure. Apart from counterterms 
$40$ diagrams  contribute to the two-loop AMM of the fermion.
After performing the Dirac and Lorentz algebra, all diagrams can be
reduced to some set of scalar prototypes. A ``prototype'' defines the
arrangement of massive and massless lines in a diagram. Individual
integrals are specified by the powers of the denominators, called indices of
the lines. From the point of view of the asymptotic expansion method
the topology of the diagram is essential. All diagrams of the toy-model
which contribute to the two-loop AMM can be
classified in terms of $9$ prototypes (we omit the pure QED diagrams).
These prototypes and their corresponding subgraphs $\gamma$ 
involved in the asymptotic expansion, are given in Fig. \ref{fig3}.
In dimensional regularization, the last subgraphs  vanish  
in cases $1, 4, 7$, and $8$, owing to massless loops. 

All diagrams were generated in symbolic form by means of QGRAF \cite{QGRAF}.
The TLAMM-program

$1.$
Reads QGRAF output;

$2.$
For each diagram, it creates a file containing  the complete 
FORM program for calculating this diagram;

$3.$
Executes FORM;

$4.$
Reads FORM output, picks out the result of the 
calculation, and builds the total sum of all diagrams in a single file.

\noindent
All initial settings are defined in the ``configuration file''. It contains 
information about the use of file names, identifiers of topologies, 
choice of momenta, and the description of the model in terms of the 
notation that is some extension of QGRAF's. 
All diagrams are classified according to their prototypes.
Identifiers for vertices and propagators and explicit Feynman rules 
are read from separate files and  then inserted into the FORM program.
For each diagram, the relevant FORM subroutines are called.
Since the number of identifiers needed for the calculation of all
diagrams may exceed FORM capacity, the TLAMM-program  keeps for each 
diagram only those needed for its calculation. Finally,
the total sum of all diagrams is put into one file and can be processed 
further by FORM. There exist several options which allow one to process 
only diagrams
$1$ with given numbers;
$2$ of a given prototype;
$3$ of a specified topology;
and some debugging options.

To perform the asymptotic expansion FORM-programs have been written
for every prototype. 
For efficiency of the algorithm the following is essential:

$1.$
The result of the calculation is presented as a series in small
parameters. Care is taken to avoid the production of unnecessary high
powers in intermediate results.

$2.$
For the evaluation of the Feynman integrals it is necessary to reduce scalar
products of momenta in the numerator to the square combinations
present in the denominator. Most efficiently, this is done by means of 
recurrence relations proposed by Tarasov \cite{Tarasov} in the
proceeding of the AIHENP-$95$.

We use the on-shell renormalization scheme. 
The renormalization conditions are as follows:
$1.$
The electric charge $e$ is defined in terms of the  nonrelativistic 
Thompson limit of the Compton scattering.
$2.$
The physical masses are given as the pole of the propagators.
Since we are interested here only in the AMM, we don't need the
wave-function renormalization. All the tadpoles are canceled 
by mass counterterms. Then at the two-loop
level the interaction $\Phi^4$ does not contribute to the AMM of the 
fermion. No special functions but the logarithms appear in the final 
result:

\begin{eqnarray}
a_\mu & = & 
\frac{e^2}{4 \pi^2} \left\{\frac{1}{2}\right\}
+ \frac{e^4}{16 \pi^4}\left\{
\frac{197}{144} +\left( \frac{1}{2}-3 \ln \left(2\right)\right) 
\zeta \left(2 \right) +\frac{3}{4} \zeta \left(3 \right) \right\}
\nonumber \\[3mm]
&& 
+ \frac{y^2}{16 \pi^2}
\left(\frac{m}{M}\right)^2\left\{  
- \frac{7}{3}+2\ln\left(\frac{M}{m}\right)^2
\right\}
\nonumber \\[3mm]
&&
+\frac{e^2y^2}{64 \pi^4}
\left(\frac{m}{M}\right)^2\left\{  
\frac{335}{27}+\frac{121}{9}\ln\left(\frac{M}{m}\right)^2
-\frac{13}{2}\ln^2\left(\frac{M}{m}\right)^2
-29 \zeta \left(2 \right)
\right\}
\nonumber \\[3mm]
&&
+\frac{y^3 }{256 \pi^4 } \left(\frac{g m}{M^2}\right)
\left\{ 2  - \frac{4}{3} \zeta \left(2 \right)
 + \left(\frac{m}{M}\right)^2
\left\{\frac{46}{3}-\frac{189}{2}S_2
+6\ln\left(\frac{M}{m}\right)^2
\right\} \right\}
\nonumber \\[3mm]
&&
+\frac{y^2}{256 \pi^4}
\left(\frac{g m}{M^2}\right)^2\left\{  
- \frac{5}{3}+\frac{45}{2}S_2
-\frac{13}{3}\frac{\pi}{\sqrt{3}}
-\left(2-\frac{16 \pi}{\sqrt{3}}\right)
\ln\left(\frac{M}{m}\right)^2
\right\}
\nonumber \\[3mm]
&& 
+\frac{y^4}{256 \pi^4}
\left(\frac{m}{M}\right)^2\left\{  
\frac{103}{6}-13\ln\left(\frac{M}{m}\right)^2
\right\}
\end{eqnarray} 
where 
$S_2 = \frac{4}{9 \sqrt{3}} {\it Cl}_2 
\left( \frac{\pi}{3} \right) = 0.2604341$
with ${\it Cl}_2$ the  Clausen  function, 
$e^2/(4 \pi) \equiv \frac{1}{137}$ and $y, g$ are
renormalized coupling constants. We see that

$1$ 
The two-loop contribution to the AMM is gauge independent.

$2$
The additional threshold singularities arising in the 
asymptotic R-operation have canceled. 

$3$
With the Standard Model values it is 
$ \sim \left( \frac{m}{M} \right) ^4$.

\nonumsection{References}

\end{document}